\documentclass[a4paper, fontsize=11pt, oneside, twocolumn=false, openright,
parskip=half-, titlepage=false, cleardoublepage=empty,
bibliography=totoc, BCOR=5mm, DIV=12]{scrartcl}

\usepackage[utf8]{inputenc}
\usepackage[T1]{fontenc}
\usepackage[titletoc]{appendix}
\usepackage{mathtools,xfrac,amsfonts,amssymb,braket,physics,bbm,mathtools,xcolor,mathrsfs}
\usepackage[multiple]{footmisc}
\usepackage{dsfont}
\usepackage{makeidx}
\usepackage{graphicx}
\usepackage{cite}
\usepackage{float}
\usepackage{subcaption}
\usepackage[font=footnotesize]{caption}

\usepackage{authblk}

\title{Coarse-grained dynamics of ac-driven two-state systems}

\author[2,3]{Nicola Macrì}
\author[1,2]{Luigi Giannelli}
\author[2,3]{Jishnu Rajendran}
\author[1,2,3]{Elisabetta Paladino}
\author[1,2,3]{Giuseppe Falci}

\affil[1]{CNR-IMM, UoS Università, 95123 Catania, Italy}

\affil[2]{Dipartimento di Fisica e Astronomia ``Ettore Majorana'', Universitá di
	Catania, 95123 Catania, Italy}

\affil[3]{INFN Sezione di Catania, 95123 Catania, Italy}

\date{}

\begin{document}
	\maketitle
	\begin{abstract}
		Magnus expansion is used to identify effective Hamiltonians describing the coarse-grained dynamics of more complex problems. Here, we apply this method to a two-level system driven by an and AC field. We derive Stark and Bloch-Siegert shifs of both diagonal and off-diagonal entries of the Hamiltonian as a result of coarse-graining only. 
	\end{abstract}
	\section{Introduction: effective Hamiltonian by Magnus Expansion}
	The Magnus expansion is a mathematical tool allowing one to express the solution of a linear differential equation as an exponentiated series. In particular the time-evolution operator $U(t, t_0)= -\mathrm{e}^{ -i \, F(t,t_0) }$ solving the Schr\"odinger equation $ \dot{U}(t,t_0) = - i \, H(t) \, U(t,t_0) $ with initial condition $ U(t_0,t_0) = \mathds{1} $ is expressed as a series $F(t,t_0) = \sum_i F_i(t,t_0)$ where (higher-order terms can be found in Ref. \cite{BLANES2009151})
	\begin{equation}
	\label{eq:magnus-terms}
	F_1 = \int_{t_0}^t \, ds \, H(s),
	\quad ; \quad 
	F_2 = -\frac{i}
	{2} \int_{t_0}^t \int_{t_0}^{s_1} ds_1 \, ds_2 \, \comm{H(s_1)}{H(s_2)}
	\end{equation}
	The small expansion parameter is related to $[H(t),H(t^\prime)]$. The term $F_1$ yields the average Hamiltonian \cite{RevModPhys.76.1037}, while $F_2$ provides already an excellent approximation in many cases.\\ 
	Coarse graining over a time $\tau$ is obtained by approximating $U(t+\tau/2,t-\tau/2) \approx \mathrm{e}^{-i H_\mathrm{eff}(t) \tau }$, where the effective Hamiltonian is obtained by truncating the Magnus series,  $H_\mathrm{eff}(t) = \tfrac{1}{\tau} \, \left[F_1 + F_2 + \ldots \right]$. The  coarse grained dynamics is then expressed as $U_{eff}(t, t_0) = \mathrm{T} \, \mathrm{exp} \Big\{- i \int_{t_0}^t ds \, H_\mathrm{eff}(s) \Big\}$. 	If $[H(t^\prime),H(t^{\prime\prime})]$ is sufficently small for $ t^{\prime\prime} \neq t' \in [t+\tau/2,t-\tau/2]$, the effective Hamiltonian involves few terms of the series and it is expected to be simpler than the original $H(t)$.
	
	We validate the approximation via the fidelity ${\cal F}(t) = \min_{ \ket{\psi_0} } \big( \big|\bra{\psi_0} U^{\dagger}(t) \, U_\mathrm{eff}(t) \ket{\psi_0}\big|^2 \big) $, which is a simple unitarily invariant figure of merit directly referring to the dynamics. The analytical expression ${\cal F} = \Tr{ U^{\dagger} \, U_\mathrm{eff} } $ holds for a two-state system.
	\section{Results for ac driven 2-level atom}
	We consider a two-level system with Hamiltonian $H_0 = - \frac{ \epsilon }{ 2 } \, \hat{\sigma}_3$ driven by a monochromatic ac field $H_1 = H_{RW} + H_{CR}$ with corotating $H_{RW}= {\mathscr W }\, \mathrm{e}^{ i \, \omega t } \sigma^+ + \mathrm{h.c.} $ and counterrotating $H_{CR}= {\mathscr W }\, \mathrm{e}^{ -i \, \omega t } \sigma^+ + \mathrm{h.c.} $ terms. In the interaction picture, $\tilde H(t) = \mathrm{e}^{i H_0 t} H_1 \mathrm{e}^{-i H_0 t} $ the two lowest order terms of the Magnus expansion are calculated
	\begin{equation}
	\label{eq:detuned}
		\begin{aligned}
		&\tilde{H}_{eff}^{(1)} = \frac{\mathscr W }{ 2 } \Big( \, \mathrm{e}^{-i \, \delta t } \, \mathrm{sinc} \frac{ \delta \tau }{ 2 } + \mathrm{e}^{-i \, (\epsilon + \omega) t } \, \mathrm{sinc}  \frac{(\epsilon + \omega) \tau }{ 2 } \, \Big) \, \hat{\sigma}^- + \mathrm{h.c} \\
		&\begin{split}
		\tilde{H}_{eff}^{(2)} = & \Big\{ -\frac{ \mathscr W^2 }{ 4 } \Big[ \frac{ 1 - \mathrm{sinc} \,\delta \, \tau  }{ \delta }  + \frac{ 1 - \mathrm{sinc} ( \epsilon + \omega ) \, \tau }{ \omega + \epsilon }\Big] - \frac{ \mathscr W^2 }{ 4 } \, \mathrm{e}^{ \, 2 i \, \omega t} \, \Big[\frac{\, \mathrm{sinc} \, \omega \tau}{\delta (\omega + \epsilon) } \\ 
		&-\Big( \frac{ \mathrm{sinc} (\delta \tau /2) }{ \omega + \epsilon } \, \mathrm{e}^{-i ( \epsilon + \omega ) \, \tau /2 } + \frac{ \mathrm{sinc} [( \epsilon + \omega ) \tau/ 2 ] }{ \delta } \,  \, \mathrm{e}^{i \delta \, \tau /2 } \Big) \Big] + \mathrm{c.c.} \Big\} \hat{\sigma}_3 \nonumber
		\end{split}
		\end{aligned}
	\end{equation}
	where $\delta= \epsilon-\omega$ is the detuning. Since 
	$ \omega, \epsilon > 0 $ we can eliminate the fast dynamics by choosing $ \tau \gg {\pi \over \omega}, \frac{ 2 \pi }{ \omega + \varepsilon }$, meaning that the $ \mathrm{sinc}(x) $ terms are very small except possibly those containing $\delta$. These latter can also be neglected if $\delta \tau \gg 2 \pi$. In this case $\tilde{H}_{eff}^{(1)} \approx 0$ while $\tilde{H}_{eff}^{(2)} \approx - {1 \over 2} [S_{RW} + S_{BS}] \,\sigma_3$, where the Stark shift 
	$S_{RW} = |\mathscr W |^2/( 2\delta)$ and the diagonal Bloch-Siegert shift $S_{BS} = |\mathscr W |^2/2 \, ( 2 \omega + \delta )$ \cite{allen1987optical} appear. 
	This result is self-consistent if we can choose $\tau \ll 2 \pi/S_{RW}$, such that the effect of the shifts on top of the bare dynamics is apparent. Thus the approximation holds only if $\delta / \mathscr W \gg 1$, \textit{i.e.} for large detuning.
	
\begin{figure}[t]
    \begin{subfigure}{.5\textwidth}
        \centering
        \includegraphics[width=1.0\linewidth]{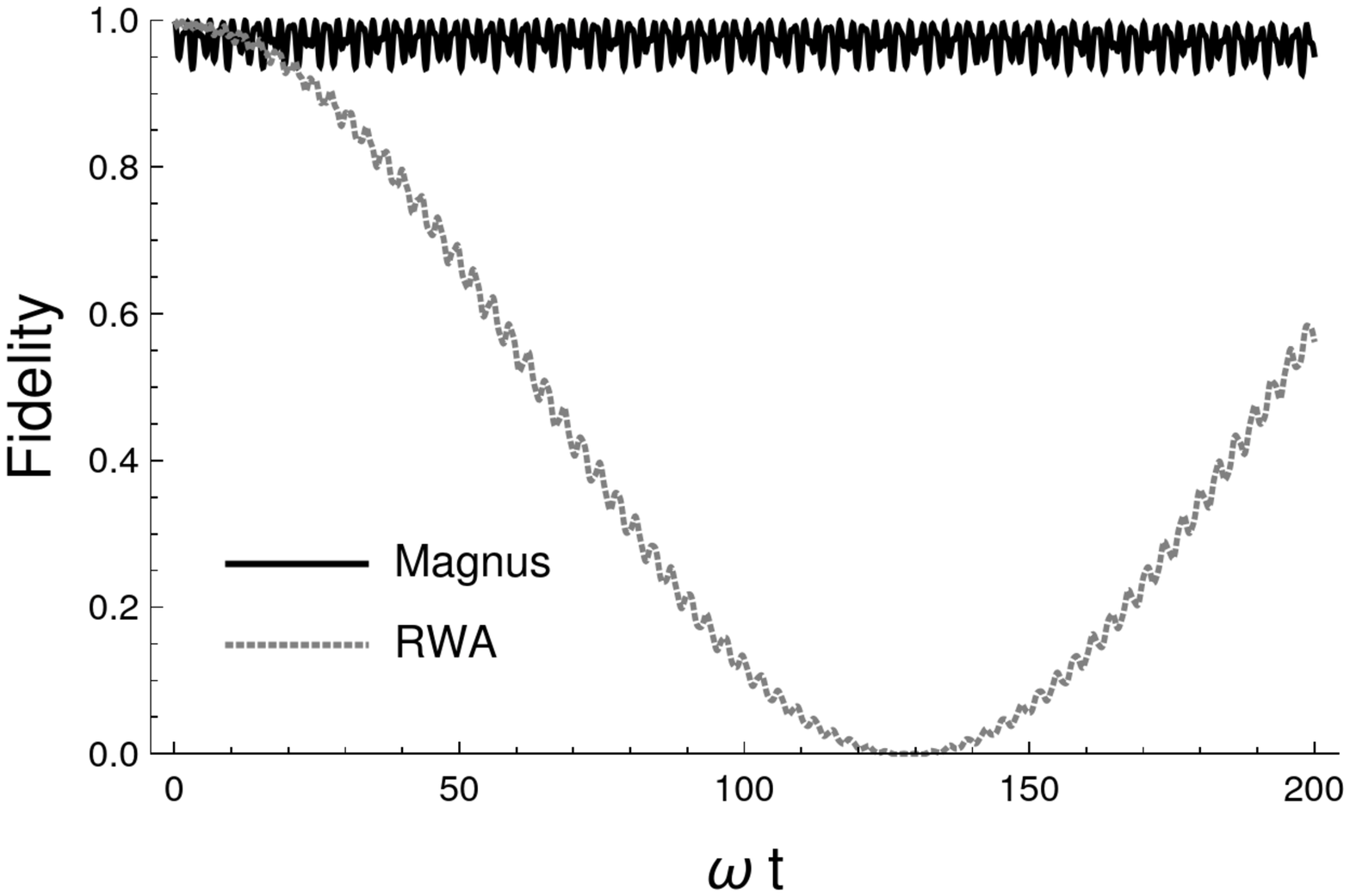}
        \caption{}
    \end{subfigure}
    \begin{subfigure}{.5\textwidth}
        \centering
        \includegraphics[width=1.\linewidth]{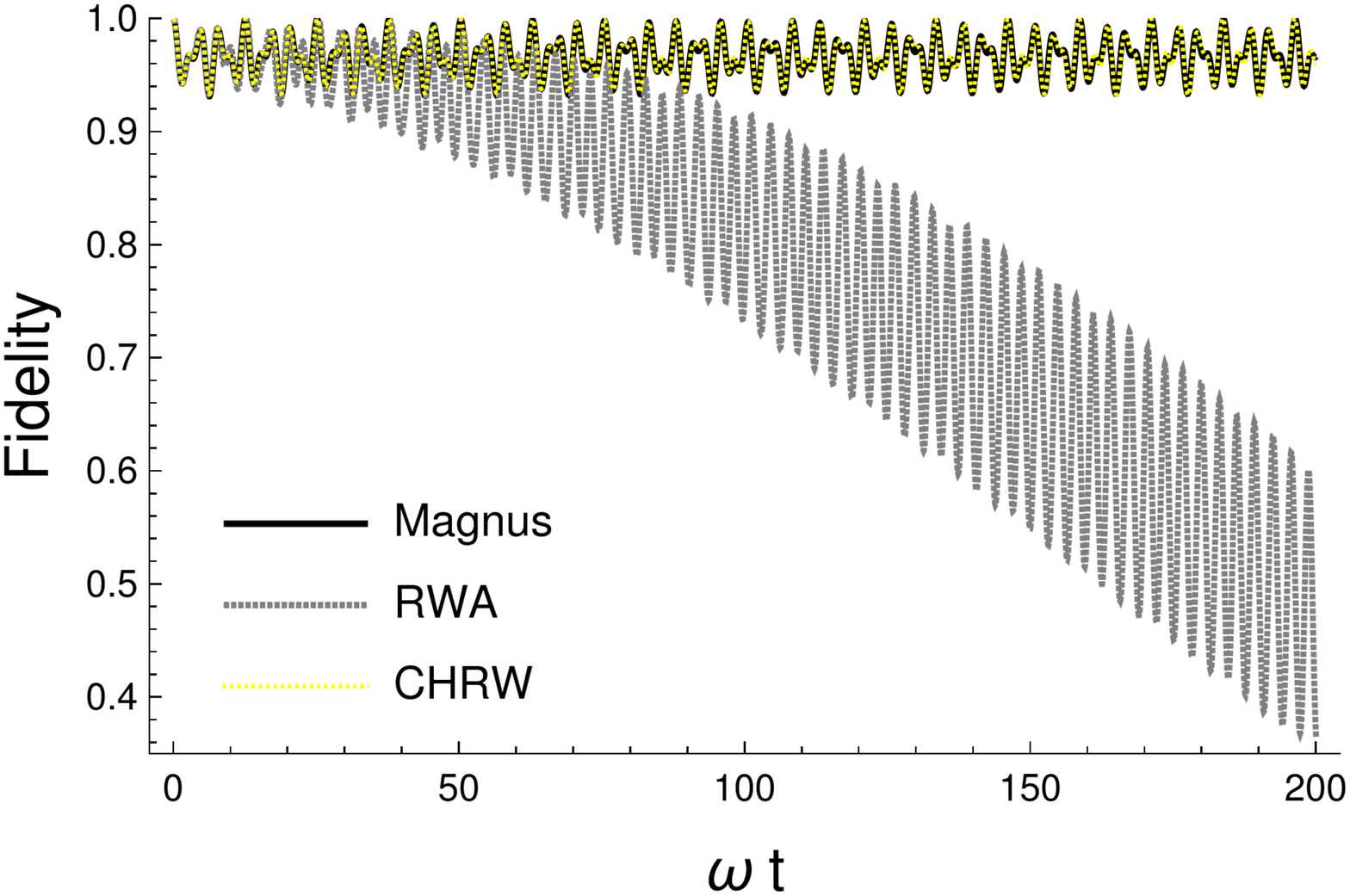}
        \caption{}
    \end{subfigure}
    \caption{ Fidelities for the Magnus effective Hamiltonian at second-order for $\mathscr W/\omega = 0.5$. (a) For $\delta/\omega = 3$ and compared to the RWA. (b) At resonance, compared with the RWA plus the diagonal $S_{BS}$ and with the CHRW method~\cite{PhysRevA.86.023831, PhysRevA.75.063414}.}
    \label{fig:fids_RWA_Magn}
    \end{figure}

	The dynamics determined by $H_{eff}$ conserves populations of the eigenstates of $\sigma_3$ accumulating a dynamical phase between them. We validate this result by the fidelity (see Fig.\ref{fig:fids_RWA_Magn}a) with respect to the exact dynamics, evaluated up to relatively large times which amplify the effect of errors in the dynamical phase. Values of the parameters are chosen in order to test the limits of our approximation, as the quite large $\mathscr W/\omega= 0.5$ and not so large 
	$\delta/\omega=3$. It is seen that ${\cal F} \gtrsim 0.95$, the error being due to small changes in the populatiions. We also plot the fidelity for the rotating wave approximation (RWA) $H=H_{RW}$ showing that the diagonal $S_{BS}$ cannot be neglected. 
	
	Notice that if we let $\delta \to 0$ in Eq.(\ref{eq:detuned}) we would obtain $\tilde{H}_{eff}^{(1)} \to \tilde H_{RW}$ while  in $\tilde{H}_{eff}^{(2)}$ terms containing $\delta$ vanish. Therefore it is tempting to take
    ${H}_{eff}= - {1 \over 2}\,S_{BS}\, \sigma_3 + H_{RW}$ as often done in the standard treatment of quantum optics~\cite{allen1987optical}. However this guess is not always accurate: in fact at resonance a new term appears which renormalizes $\mathscr W$. The calculation is carried in a rotating frame defined by the transformation $U_x(t) = \mathrm{e}^{-i H_0 t} \mathrm{e}^{-i \tilde H_{RW} t}$. In this frame the Hamiltonian is $\bar H(t) =U^\dagger_x(t) \, \tilde{H}_{CR}(t)\, U_x(t)$ and we use it to evaluate the first two terms of the Magnus expansion. Letting $\omega > \mathscr W > 0$ we choose $  \frac{ 2 \pi }{ \mathscr W } \gg \tau \gg  \frac{2 \pi}{ 2 \, \omega - \mathscr W } $. Thus, the first-order term averages out while at second order slowly varying terms are retained, obtaining
    $$
    \bar H_{eff}^{(2)} = -  {S_{BS}^\prime \over 2}\, \sigma_1 -  {S_{BS} \over 2}\, {1 - (\mathscr W/2\sqrt{2} \omega)^2
     \over 1- (\mathscr W/2 \omega)^2   } \, \big[ \mathrm{e}^{- i \mathscr W t}\,\ketbra{+}{-} + \mathrm{h.c.} \big]
    $$
    where $\ket{\pm}$ are eigenstates of $\sigma_1$ and the off-diagonal Bloch-Siegert shift is defined as
    $$ S^\prime_{BS} =
    {\mathscr W^3 \over 16 \, \omega^2 [1- (\mathscr W/2 \omega)^2 ]  }
    $$
     This result is self-consistent only if we can choose $\tau \ll 2 \pi/S_{RW}^\prime$ which implies that it must be $\mathscr W/\omega \gg (32)^{-1/3}$. 
    Transforming back to the interaction picture we recover the more familiar structure
    $$
    \tilde{H}_{eff}= -  {1 \over 2}\,S_{BS}\, \sigma_3 +   {1 \over 2}\, (\mathscr 
    W - S_{BS}^\prime) \sigma_1 
    $$
    Contrary to what previously guessed this result shows that the field amplitude is renormalized by the "off-diagonal" shift $S_{BS}^\prime$.
	The fidelity for our $H_{eff}$ and for the RWA plus diagonal $S_{BS}$  guess which neglects the "off-diagonal" $S_{BS}^\prime$ is shown in Fig.\ref{fig:fids_RWA_Magn}b.  We choose $\mathscr W/\omega= 0.5$ which is demanding for our approximation. It is clear that in this regime $S_{BS}^\prime$ cannot be neglected. This may be surprising at first sight since  in the regime $W/\omega < 1$ we consider $S_{BS} \sim \mathscr W^2/\omega$ is larger than $S_{BS}^\prime \sim \mathscr W^3/\omega^2$, but it can be understood by noticing that both shifts appear at the same order of $ \mathscr W/\omega$ in the eigenvalues of $\tilde{H}_{eff}$.
	\section{Conclusion}
	We studied the dynamics of a two-state system driven by a monochromatic ac classical field with both corotating and counterrotating components. This problem has been tackled in the past by several accurate and elaborated analytic methods. We show that a satisfactory and analytically solvable effective Hamiltonian can be found by deriving the coarse grained dynamics, with no further basic assumption. Coarse graining was operated approximating the dynamics in short time intervals via the Magnus expansion. Our results agree with those obtained by more elaborated methods~\cite{ PhysRevA.86.023831, PhysRevA.75.063414}, as also shown in Fig.\ref{fig:fids_RWA_Magn} for the  CHRW method of Ref.\cite{PhysRevA.86.023831, PhysRevA.75.063414} at first order in $ \mathscr W/\omega $. The advantage of our approach is that it relies only on coarse-graining and it can be extended to more general instances, as slowly varying field amplitudes and frequency chirping~\cite{FalciFP2017advances}, multitone drives and multilevel systems~\cite{DiStefanoPRA2016coherent,DiStefanoPRB2015populationa}, operations in devices with realistic noise~\cite{PaladinoRMP2014oneoverfnoise}. In view of the accuracy we obtain for relatively large values of $\mathscr W$ we expect that extensions of our method to quantum fields may provide accurate effective Hamiltonians for atom-cavity systems in the ultrastrong coupling regime~\cite{ka:20-pellegrino-commphys-1overfgraph,FalciSR2019ultrastrong}, which are a subject of large interest in Condensed Matter physics and Quantum Technologies.

    \section*{Acknowledgments} This work was supported by the QuantERA grant SiUCs (Grant No.731473), and by University of Catania, Piano Incentivi Ricerca di Ateneo 2020-22, progtto Q-ICT.

	\bibliographystyle{plain}
	\bibliography{bib}

\begin{thebibliography}{10}

\bibitem{allen1987optical}
Leslie Allen and Joseph~H Eberly.
\newblock {\em Optical resonance and two-level atoms}, volume~28.
\newblock Courier Corporation, 1987.

\bibitem{PhysRevA.75.063414}
S.~Ashhab, J.~R. Johansson, A.~M. Zagoskin, and Franco Nori.
\newblock Two-level systems driven by large-amplitude fields.
\newblock {\em Phys. Rev. A}, 75:063414, Jun 2007.

\bibitem{BLANES2009151}
S.~Blanes, F.~Casas, J.A. Oteo, and J.~Ros.
\newblock The magnus expansion and some of its applications.
\newblock {\em Physics Reports}, 470(5):151--238, 2009.

\bibitem{DiStefanoPRB2015populationa}
P.~G. Di~Stefano, E.~Paladino, A.~D'Arrigo, and G.~Falci.
\newblock Population transfer in a {{Lambda}} system induced by detunings.
\newblock {\em Phys. Rev. B}, 91(22):224506, June 2015.

\bibitem{DiStefanoPRA2016coherent}
P.~G. Di~Stefano, E.~Paladino, T.~J. Pope, and G.~Falci.
\newblock Coherent manipulation of noise-protected superconducting artificial
  atoms in the {{Lambda}} scheme.
\newblock {\em Phys. Rev. A}, 93(5):051801, May 2016.

\bibitem{FalciFP2017advances}
G.~Falci, P.~G. Di~Stefano, A.~Ridolfo, A.~D'Arrigo, G.~S. Paraoanu, and
  E.~Paladino.
\newblock Advances in quantum control of three-level superconducting circuit
  architectures.
\newblock {\em Fortschritte der Physik}, 65(6-8):1600077, 2017.

\bibitem{FalciSR2019ultrastrong}
G.~Falci, A.~Ridolfo, P.~G. Di~Stefano, and E.~Paladino.
\newblock Ultrastrong coupling probed by {{Coherent Population Transfer}}.
\newblock {\em Sci Rep}, 9(1):9249, June 2019.

\bibitem{PhysRevA.86.023831}
Zhiguo L\"u and Hang Zheng.
\newblock Effects of counter-rotating interaction on driven tunneling dynamics:
  Coherent destruction of tunneling and bloch-siegert shift.
\newblock {\em Phys. Rev. A}, 86:023831, Aug 2012.

\bibitem{PaladinoRMP2014oneoverfnoise}
E.~Paladino, Y.~M. Galperin, G.~Falci, and B.~L. Altshuler.
\newblock \$\textbackslash mathbsf\{1\}/\textbackslash mathbsfit\{f\}\$ noise:
  {{Implications}} for solid-state quantum information.
\newblock {\em Rev. Mod. Phys.}, 86(2):361--418, April 2014.

\bibitem{ka:20-pellegrino-commphys-1overfgraph}
M.~D.~Francesco Pellegrino, Giuseppe Falci, and Elisabetta Paladino.
\newblock 1/f critical current noise in short ballistic graphene josephson
  junctions.
\newblock {\em Comm. Phys.}, 3:6, 2020.

\bibitem{RevModPhys.76.1037}
L.~M.~K. Vandersypen and I.~L. Chuang.
\newblock Nmr techniques for quantum control and computation.
\newblock {\em Rev. Mod. Phys.}, 76:1037--1069, Jan 2005.

\end{thebibliography}
	
	\end{document}